**Title:** Facial expressions can detect Parkinson's disease: preliminary evidence from videos collected online.

**Authors:**

Mohammad Rafayet Ali, Ph.D.
Postdoctoral Associate
Computer Science, University of Rochester
Email: mali7@cs.rochester.edu

Taylor Myers, B.A.
Research Coordinator
Center for Health + Technology
University of Rochester
Email: Taylor.Myers@chet.rochester.edu

Ellen Wagner, M.S.
UX Specialist
Center for Health + Technology
University of Rochester
Email: Ellen.Wagner@chet.rochester.edu

Harshil Ratnu
Undergraduate Researcher
Computer Science, University of Rochester
Email: hratnu@u.rochester.edu

E. Ray Dorsey, M.D
Professor
Department of Neurology, Center for Health + Technology
University of Rochester
Email: ray.dorsey@chet.rochester.edu

Ehsan Hoque, Ph.D.
Associate Professor
Computer Science, University of Rochester
Email: mehoque@cs.rochester.edu

**Abstract**

One of the symptoms of Parkinson's disease (PD) is hypomimia or reduced facial expressions. In this paper, we present a digital biomarker for PD that utilizes the study of micro-expressions. We analyzed the facial action units (AU) from 1812 videos of 604 individuals (61 with PD and 543 without PD, mean age 63.9 y/o, sd 7.8 ) collected online using a web-based tool (www.parktest.net). In these videos, participants were asked to make three facial expressions (a smiling, disgusted, and surprised face) followed by a neutral face. Using techniques from computer vision and machine learning, we objectively measured the variance of the facial muscle movements and used it to distinguish between individuals with and without PD. The prediction accuracy using the facial micro-expressions was comparable to those methodologies that utilize motor symptoms. Logistic regression analysis revealed that participants with PD had less variance in AU6 (cheek raiser), AU12 (lip corner puller), and AU4 (brow lowerer) than non-PD individuals. An automated classifier using Support Vector Machine was trained on the variances and achieved 95.6% accuracy. Using facial expressions as a biomarker for PD could be potentially transformative for patients in need of physical separation (e.g., due to COVID) or are immobile.

1. Introduction

Parkinson's disease can cause tremors in limbs, muscular rigidity, non-rhythmic head movements, soft speech, and much more [1]. One of the major symptoms of PD is stiffness in facial muscle movements and reduced facial expression [2]. This phenomenon is known as hypomimia. Hypomimia is considered an extremely sensitive biomarker for PD [3] making it a very useful tool for early diagnosis of PD. It is very important to detect PD early since the reduced facial expressions are linked to reduced social wellbeing and depression [4].

The existing digital biomarkers which capitalize on the motor symptoms often use wearable sensors (e.g., MC10 BioStampRC sensor[5], tri-axial gyroscope[6]) that are reliable but require active participation, are expensive and thus, not scalable. On the other hand, facial expression analysis can be used as a non-invasive tool for detecting PD which can be done using a webcam or a phone with a camera. An additional benefit of analyzing videos is that potential PD patients do not need to be in close proximity to a neurologist for an in-person diagnosis. This is potentially transformative for patients in need of physical separation from others (e.g., due to COVID-19) or are immobile. In addition, many underdeveloped areas and underrepresented populations can also benefit from such digital biomarkers without worrying about direct access to a neurologist.

In this work, we analyzed 1812 videos collected using an online tool via the Internet (available at www.parktest.net). The videos contain 543 individuals performing three facial mimicry tasks. The facial mimicry tasks include making three facial expressions three times followed by a neutral face after every expression. The three facial expressions were smiling face, disgusted face, and surprise face. We analyzed the facial muscle movements using the facial action coding system [7] while the participants were showing a facial expression. We computed the variance of the muscle movements in terms of action units (AUs) and showed that this measure of micro-expressions has the potential to be an important digital biomarker for PD. Our analysis shows that individuals with PD have fewer facial muscle movements than non PD individuals (i.e., low variance in AU4 Brow Lowerer, AU6 Cheek Raiser, and AU12 Lip Corner Puller). We also find that the PD prediction accuracy using the facial muscles (95%) from videos is comparable to the existing state of the art video analysis that utilizes limb movements (~93%) [8].

2. Results

Table 1 shows the demographics information. Table 2 shows the difference in the variance of AUs between PD and non PD population. The significant differences were found in the smiling and surprising faces. Specifically, AU01 (inner brow riser), and AU06 (cheek raiser) had a significant difference in smiling facial expression. This difference indicates that smiling facial expression has the greatest potential in differentiating individuals with and without PD.

We applied an SVM on the variance of nine AUs to classify individuals with PD. The result of the classifier is below.
Accuracy: 95.6%, F1: 0.95, AUC: 0.94, Precision: 95.8%, Recall: 94.3%

Figure 1 shows the weights of the features from logistic regression. The dependable variable was the binary diagnosis of PD and non PD (PD being 1 and non PD being 0). The green bars indicate those features where p < 0.05. Seven out of nine features had negative weights. This means a lower variance in AUs indicates a higher probability of having PD. This also indicates that the stiffness among facial muscle in PD participants had influenced the weights to be negative.

Figure 2 shows another visualization of the features. The nine features were converted to two-dimensional data using PCA and then the K-Means clustering was applied. Three clusters were identified. The cluster in red color had the highest percentage of PD participants (76%). Also, it should be noted that the red cluster center is close to the (0, 0) coordinate. This indicates a similar effect of variance in AUs among the PD participants.

### 3. Discussion

The distributions show that the variances of two AUs were significantly different between PD and non-PD. It is important to notice that both of these AUs were from the smiling facial expression. This indicates the smiling facial expression may be the most affected by PD among all other fundamental facial expressions. This phenomenon was also evident in past studies [9]. The weights of logistic regression (Figure 1) also show that all three of the AUs from smiling facial expression had significant weight along with one AU from a disgusted face. Three out of these four significant weights were negative indicating an opposite relation to the PD. In other words, the lower level of muscle movements (i.e., Frontalis, pars medialis, Depressor Glabellae, Depressor Supercilli, Currugator, Orbicularis oculi, pars orbitalis, and Zygomatic Major) were associated with PD.

In both Table 2 and Figure 1, AU01 has a higher level of variance in PD than non PD group. AU01 is characterized by an inner eyebrow-raiser. In the past, the eyebrow tremor was found to be an initial manifestation of PD [10]. Thus unlike other facial muscles, we think having PD was the driving factor of the higher level of eyebrow movements.

A simple SVM classifier was able to achieve 95% classification accuracy. PD symptoms are often characterized by multiple different modalities including limb tremor, head movements, voice, memory, sleep, and walking. The current state of the art video analytic tools that rely on hand tremor and head movements were able to achieve 92% accuracy [8]. Our findings provide evidence that facial expressions can be used as a reliable biomarker for PD detection.

A computer algorithm analyzing micro-expressions can further augment a neurologist's ability to objectively rate the symptoms of PD patients. A neurologist can only count the number of smiles or track the time involved in modulating a patient's expression. We show that an algorithm's ability to analyze the subtle characteristics of facial expressions, often invisible to a naked eye, adds significant new information to a neurologist.

It should be noted that not all individuals with PD show all symptoms. This means a person with PD can have hand tremors but no stiffness in facial muscles or vice versa. This is why relying on only one modality cannot achieve perfect accuracy unless, in a very unlikely event, all PD individuals in the dataset show all symptoms. Because of this variability in symptom manifestation, it is important to find reliable biomarkers that utilize multiple different modalities to develop a robust AI model for identifying PD. Our findings only show that facial expressions, especially smiling, can be used as one of the reliable modalities.

One of the most common use cases of phones is taking photos of each other. Imagine, with the user's permission, an app that can automatically analyze the characteristics of the users' smile (from a short video) and provide a referral in case they are at risk. Additionally, this can enable low-cost screening tools where access to a neurologist is limited. For example, the population/neurologist ratio in African nations far exceeds than the US and other developed countries (one study notes in 23 African countries, the average population per neurologist exceeds half a million [11]. In contrast, approximately 75% of people in Africa have access to a mobile phone. In more developed regions such as in South Africa, more than 90% of the population has access to a mobile phone [12]. A similar situation exists in Asia and South America [13]. This indicates that facial expression-based analysis could potentially enable neurologists to serve patients from underserved communities, further contributing to health equity and access.

## 4. Methods

4.1 Data Set

The dataset consists of 1812 videos from 604 (61 with PD, 543 without PD) individuals. The videos were collected using an online PD video recording tool – PARK (Parkinson's Analysis with Remote Kinetic tasks) [14]. Using this online tool users can record their videos while performing the tasks from MDS-UPDRS (Movement Disorder Society – Unified Parkinson's Disease Rating Scale). Before performing the tasks, users watch a short video of a person performing the task. In this paper, we present the analysis of the facial mimicry tasks collected using the PARK framework. The facial mimicry tasks include three separate facial expression tasks. Each video contains one facial expression. The participants were asked to make a facial expression, hold that for a couple of seconds, and then make a neutral face three times. The facial expressions are smiling face, disgusted face, and surprised face. Figure 3 shows examples of these facial expressions. The videos have 10 to 12 seconds of duration.

The participants without PD were recruited through Facebook advertisement and Amazon Mechanical Turks. These participants self-identified as not having PD. The participants with PD were recruited by emailing the PD patients who were receiving care or agreed to participate in research studies from the University of Rochester medical center. All PD participants had some degree of PD symptoms and were diagnosed by experts from the medical center.

4.2 Feature Extraction and Computational tools

The videos were analyzed using OpenFace software [15] which automatically provides the facial action unit (AU) values of each frame. The facial action units are associated with the muscle movements of the face. Activation of each facial action unit indicates the movement of a set of facial muscles [16]. For example, activation of AU 6 (also known as Cheek Raiser) indicates the movement of two facial muscles -- Orbicularis oculi and pars orbitalis at a time. The OpenFace software gives a binary activation (0 or 1) and a raw magnitude (ranging 0 to 5) of each AU for each frame of a video that contains a human face. We analyzed the variance of the raw action unit when the corresponding action unit is active (i.e., the activation value is 1). This variance is an indicator of how much facial muscle movement occurs while showing a facial expression. A facial expression can be associated with multiple action units [17]. Since each facial expression was followed by a neutral face, we looked at the values of raw AU values to find which AUs show three distinct peaks. Through this process, we identified the AUs that are associated with the three facial expressions. Each facial expression was found to be associated with three AUs. For example, AU01 (Inner Brow Raiser), AU06 (Cheek Raiser), and AU12 (Lip Corner Puller) were found to have three distinct peaks in smiling facial expression videos, thus they are associated with smiling. The disgusted face is associated with AU04 (Brow Lowerer), AU07 (Eye Lid Tightener), and AU09 (Nose Wrinkler); the surprised face is associated with AU01 (Inner Brow Raiser), AU02 (Outer Brow Raiser), and AU04 (Brow Lowerer). Table 2 shows the AUs associated with facial expressions.

We were interested in the muscle movements in those moments when the participants showed a particular facial expression. In the past, researchers have looked into the frequency and the magnitude of facial expressions [9]. In our case, the participants were asked to show a facial expression three times which makes the frequency of expressions the same for the PD group and the non-PD group. Additionally, the expressions were acted and not spontaneous, which is why analyzing the difference in AUs' magnitude between PD and non-PD did not yield any significant outcome. Since we were interested in the muscle movements, we computed the variance of raw AUs when the binary value indicated the AU is activated (i.e., binary AU equals 1). This variance becomes the feature set of our analysis.

4.3 Analysis

The first set of analysis looks at the difference in the distribution of features between PD and non-PD. We performed the Mann-Whitney U test to identify the differences. Since the data was not normally distributed, we used the non-parametric significance test. For repeated significance test we used a Bonferroni correction to all the p-values reported here [18].

Next, we trained a machine learning model, SVM (Support Vector Machine) on the features collected from the videos to classify PD and non-PD. Our goal was to find out whether the data has any predictive power which could be utilized in the future for classification with more sophisticated machine learning models. The dataset was imbalanced which is why we used the SMOTE [19] technique to resample the data making the dataset balanced. We then used a leave-one-out cross-validation method to report the classification metrics. In this classification task, we computed the F1-score, precision, recall, and area under the curve metrics.

To find which features were important in predicting PD and non-PD, we applied a logistic regression model. We performed the Wald test to find the significance of the features as well as the weights of the fitted model.

We then focused on visualizing the data. Since the data had nine dimensions (from nine relevant AUs) we first reduced the dimension to two. We applied the principal component analysis (PCA) to reduce the dimensionality of the data. We then applied k-means clustering on the reduced dimensional data to see if there exist regions with a high percentage of PD populations.

## 5. Data Availability

Due to the IRB requirements, the videos, which contain identifiable information cannot be shared. The extracted features which are used in this analysis, as well as the python scripts, can be found here
https://github.com/mali7/PARK_facial_mimic

**Acknowledgments**

This research is funded by the National Institute of Neurological Disorders and Stroke of the National Institutes of Health under Award Number P50NS108676 and a Google Faculty Award.

**Author Contributions**

Mohammad Rafayet Ali, performed the analysis on the dataset. Taylor Myers and Ellen Wagner recruited participants and collected the dataset. Ellen Wagner also worked on the UX of the online tool. Harshil Ratnu worked on the online tool which was used to collect the dataset. Ehsan Hoque and Ray Dorsey are the PI of this research. They interpreted the findings. All authors contributed in writing the manuscript.

**References**
1. *Parkinson's Disease: Pathogenesis and Clinical Aspects*. Parkinson's Disease: Pathogenesis and Clinical Aspects (2018). doi:10.15586/codonpublications.parkinsonsdisease.2018
2. Rinn, W. E. The neuropsychology of facial expression: A review of the neurological and psychological mechanisms for producing facial expressions. *Psychol. Bull.* (1984). doi:10.1037/0033-2909.95.1.52
3. Novellino, F. & Salsone, M. *A Tool for Evaluating Facial Expression for Early Diagnosis of Parkinson's Disease*. (2018).
4. Gunnery, S. D., Habermann, B., Saint-Hilaire, M., Thomas, C. A. & Tickle-Degnen, L. The relationship between the experience of hypomimia and social wellbeing in people with Parkinson's disease and their care partners. *J. Parkinsons. Dis.* (2016). doi:10.3233/JPD-160782
5. Lonini, L. *et al.* Wearable sensors for Parkinson's disease: which data are worth collecting for training symptom detection models. *npj Digit. Med.* (2018). doi:10.1038/s41746-018-0071-z
6. Schlachetzki, J. C. M. *et al.* Wearable sensors objectively measure gait parameters in Parkinson's disease. *PLoS One* (2017). doi:10.1371/journal.pone.0183989
7. Ekman, P. & Friesen, W. V. *Facial Action Coding System: A Technique for the Measurement of Facial Movement*. Consulting Psychologists Press (1978).
8. Rafayet Ali, M., Hernandez, J., Ray Dorsey, E., Hoque, E. & McDuff, D. Spatio-Temporal Attention and Magnification for Classification of Parkinson's Disease from Videos Collected via the Internet. *Autom. Face Gesture Recognit. (FG 2020)* (2020).
9. Smith, M. C., Smith, M. K. & Ellgring, H. Spontaneous and posed facial expression in Parkinson's Disease. *J. Int. Neuropsychol. Soc.* (1996). doi:10.1017/S1355617700001454
10. Lopez Valdes, E. *et al.* Isolated eyebrow tremor as an initial presentation form of Parkinson's disease. *Mov. Disord.* (2016). doi:10.1002/mds.26688
11. Bower, J. H. & Zenebe, G. Neurologic services in the nations of Africa. *Neurology* (2005). doi:10.1212/01.WNL.0000150894.53961.E2
12. Silver, L. & Johnson, C. Majorities in sub-Saharan Africa own mobile phones, but smartphone adoption is modest. (2019).
13. Tan, C.-T. Neurology in Asia. *Neurology* **84**, 623–625 (2015).
14. Langevin, R. *et al.* The PARK Framework for Automated Analysis of Parkinson's Disease Characteristics. *Proc. ACM Interactive, Mobile, Wearable Ubiquitous Technol.* (2019). doi:10.1145/3328925
15. Baltrusaitis, T., Zadeh, A., Lim, Y. C. & Morency, L.-P. Openface 2.0: Facial behavior analysis toolkit. in *2018 13th IEEE International Conference on Automatic Face & Gesture Recognition (FG 2018)* 59–66 (2018).
16. Ekman, Paul and Friesen, Wallace V and Hager, J. C. Facial action coding system (FACS). *A Tech. Meas. facial action. Consult. Palo Alto* **22**, (1978).
17. Du, S., Tao, Y. & Martinez, A. M. Compound facial expressions of emotion. *Proc. Natl. Acad. Sci.* **2014**, (2014).
18. Armstrong, R. A. When to use the Bonferroni correction. *Ophthalmic & physiological optics : the journal of the British College of Ophthalmic Opticians (Optometrists)* (2014). doi:10.1111/opo.12131
19. Chawla, N. V., Bowyer, K. W., Hall, L. O. & Kegelmeyer, W. P. SMOTE: Synthetic minority over-sampling technique. *J. Artif. Intell. Res.* (2002). doi:10.1613/jair.953

**Table 1:** Demographic Information

|  | Individuals with PD | Individuals without PD |
|---|---|---|
| n | 61 | 543 |
| Age (mean/sd) | 68.03 / 7.1 | 63.50 / 5.4 |
| Gender (female/male) | 22/39 | 339/204 |
| Race (White/other) | 59/2 | 484/59 |
| Country(US/other) | 55/6 | 514/29 |

**Table 2**. The difference in the variance of facial action units (AUs) between individuals with and without Parkinson's disease. The variance of AUs is an indicator of facial muscle movement.

| Expression | Facial Action Unit | Individuals with Parkinson's disease Mean (SD) | Individuals without Parkinson's disease Mean (SD) | p-value |
|---|---|---|---|---|
| Smiling Face | **AU01 (Inner Brow Raiser)** | **0.15 (0.18)** | **0.07 (0.12)** | **0.001** |
|  | **AU06 (Cheek Raiser)** | **017 (0.15)** | **0.25 (0.25)** | **0.047** |
|  | AU12 (Lip Corner Puller) | 0.21 (0.18) | 0.27 (0.24) | 0.065 |
| Disgusted Face | AU04 (Brow Lowerer) | 0.19 (0.20) | 0.26 (0.31) | 0.063 |
|  | AU07 (Eye Lid Tightener) | 0.19 (0.20) | 0.24 (0.27) | 0.103 |
|  | AU09 (Nose Wrinkler) | 0.04 (0.06) | 0.04 (0.07) | 0.267 |
| Surprising Face | AU01 (Inner Brow Raiser) | 0.28 (0.28) | 0.27 (0.32) | 0.172 |
|  | AU02 (Outer Brow Raiser) | 0.15 (0.29) | 0.12 (0.18) | 0.144 |
|  | AU04 (Brow Lowerer) | 0.31 (0.37) | 0.40 (0.43) | 0.061 |

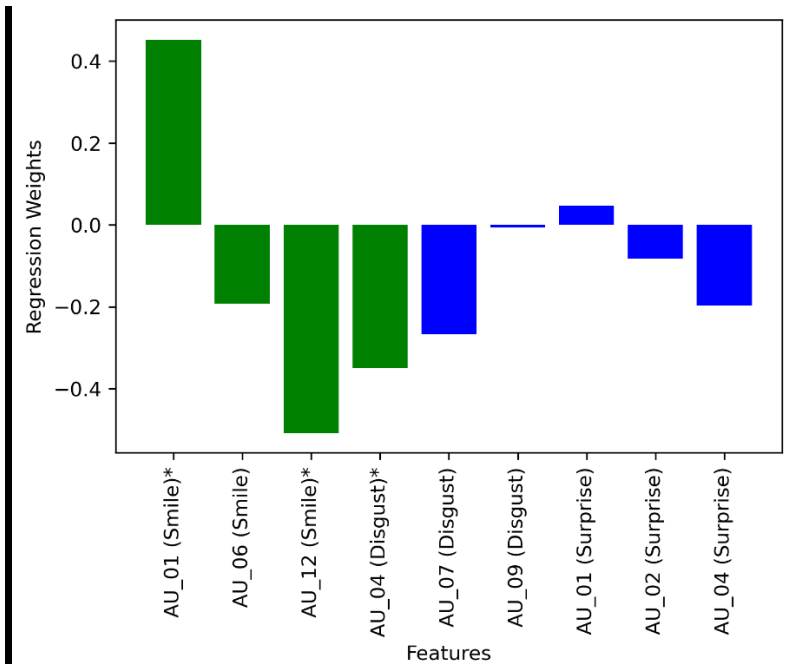

**Figure 1**: Logistic regression weights of the features while predicting PD. The green color represents the features with p<0.05.

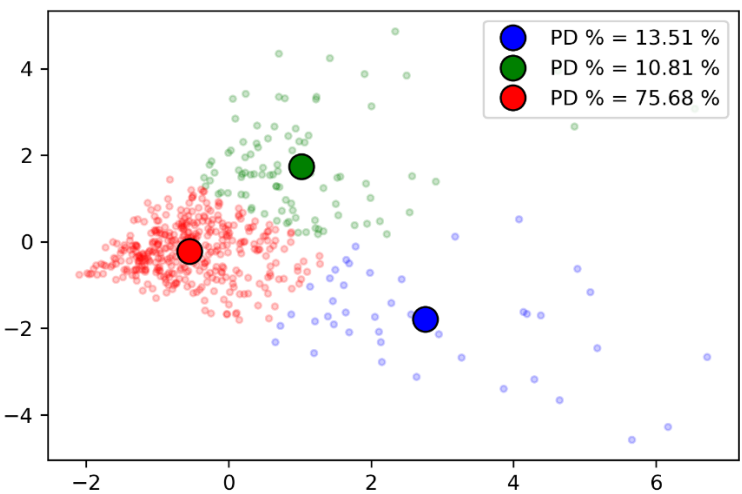

**Figure 2**: A two-dimensional visualization of the nine action units (AUs) after applying the principal component analysis (PCA). The dots on the surface represents each participant. The participants are clustered into three groups using K-means represented by three colors. The bold round dots represents the cluster center of the three clusters. The proportion of individuals with Parkinson's disease in each group differed with 75.7% of individuals in the red cluster having Parkinson's disease.

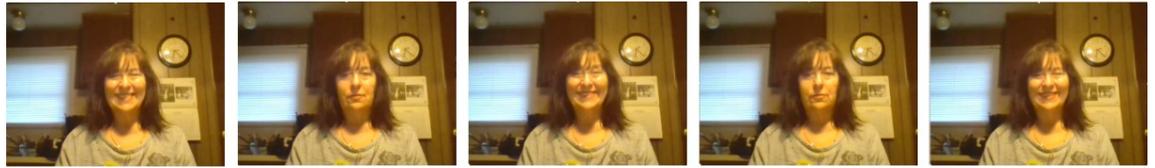
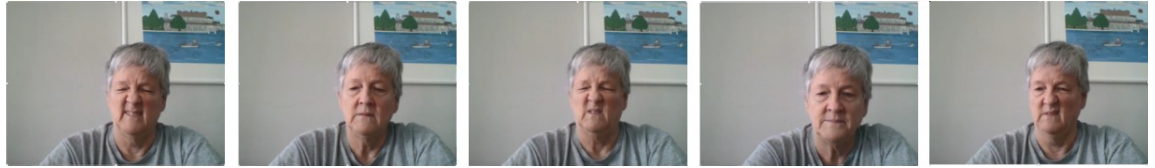
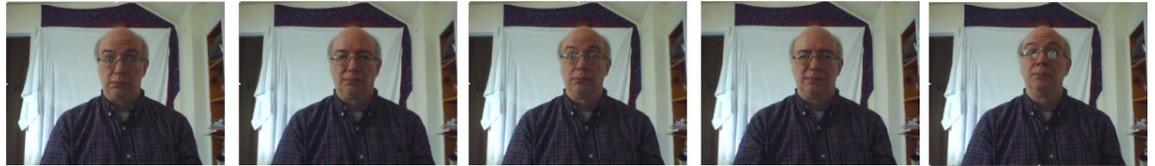

**Figure 3**. Sample frames from the data set. All participants made three facial expressions followed by a neutral face.